\documentclass[12pt]{iopart}
\usepackage{graphicx}

\usepackage{iopams}
\newcommand{\Om}{\Omega_{\rm m}}
\newcommand{\Omo}{\Omega_{\rm m}^0}
\newcommand{\Omh}{\hat{\Omega}_{\rm m}}

\newcommand{\OMo}{\Omega_{\rm M}^0}

\newcommand{\OL}{\Omega_{\Lambda}}
\newcommand{\Oc}{\Omega_{\rm c}}

\newcommand{\OLo}{\Omega_{\Lambda}^0}
\newcommand{\OX}{\Omega_{\rm X}}
\newcommand{\OXo}{\Omega_{\rm X}^0}

\newcommand{\OD}{\Omega_{\rm D}}
\newcommand{\ODo}{\Omega_{\rm D}^0}

\newcommand{\rc}{\rho_{\rm c}}
\newcommand{\rco}{\rho_{\rm c}^0}

\newcommand{\rM}{\rho_{\rm M}}
\newcommand{\rmr}{\rho_{\rm m}}

\newcommand{\rR}{\rho_{\rm R}}
\newcommand{\rD}{\rho_{\rm D}}
\newcommand{\rDo}{\rho_{\rm D}^0}
\newcommand{\rX}{\rho_{\rm X}}
\newcommand{\pX}{p_{\rm X}}
\newcommand{\wX}{\omega_{\rm X}}
\newcommand{\wm}{\omega_{\rm m}}
\newcommand{\wR}{\omega_{\rm R}}

\newcommand{\amr}{\alpha_{\rm m}}

\newcommand{\aX}{\alpha_{\rm X}}
\newcommand{\rL}{\rho_{\CC}}
\newcommand{\pL}{p_{\CC}}
\newcommand{\rLo}{\rho_{\CC}^0}
\newcommand{\pD}{p_{\rm D}}

\newcommand{\CC}{\Lambda}

\newcommand{\we}{\omega_{\rm e}}

\newcommand{\tOD}{\tilde{\Omega}_{\rm D}}

\newcommand{\lu}{\lambda_1}
\newcommand{\ld}{\lambda_2}

\newcommand{\tOmo}{\tilde{\Omega}_{\rm m}^0}

\newcommand{\rDt}{\tilde{\rho}_{\rm D}}
\newcommand{\MP}{M_{\rm P}}
\newcommand{\drmr}{\dot{\rho}_{\rm m}}
\newcommand{\drX}{\dot{\rho}_{\rm X}}

\newcommand{\rN}{r_{\rm N}}
\newcommand{\zN}{z_{\rm N}}
\newcommand{\ze}{z_{\rm e}}
\newcommand{\zs}{z_{\rm s}}

\newcommand\pubblock{\rightline{\begin{tabular}{l} \pubnumber\\
     \pubdate\\ \hepnumber \end{tabular}}}
\newcommand\pubnumber{UB-ECM-PF-06/44 }
\newcommand\pubdate{January 2007 }
\newcommand\hepnumber{gr-qc/0701090}

\begin{document}
\pubblock
\title[Cosmologies with variable parameters and dynamical cosmon]
{Cosmologies with variable parameters and dynamical cosmon:
implications on the cosmic coincidence problem}

\author{Javier Grande$^1$, Joan Sol\`{a}$^{1,}$$^2$\,
%\footnote{Speaker. Talk given at IRGAC 2006, Barcelona, July 11-15 2006.}
and Hrvoje \v{S}tefan\v{c}i\'{c}$^{1,}$$^3$}

\address{$^1$ High Energy Physics Group, Dept. ECM,  Universitat de Barcelona,\\ \hspace{0.2cm}
Diagonal 647, 08028 Barcelona, Catalonia, Spain}

\address{$^2$ C.E.R. for Astrophysics, Particle Physics and
Cosmology\,\footnote{Associated with Instituto de Ciencias del
Espacio-CSIC.}, Barcelona}

\address{$^3$ Theoretical Physics Division, Rudjer Bo\v{s}kovi\'{c} Institute, P.O.B. 180,\\ \hspace{0.2cm} HR-10002 Zagreb, Croatia}

\eads{jgrande@ecm.ub.es, sola@ifae.es and shrvoje@thphys.irb.hr}

\begin{abstract}
Dynamical dark energy (DE) has been proposed to explain various
aspects of the cosmological constant (CC) problem(s). For example,
it is very difficult to accept that a strictly constant
$\Lambda$-term constitutes the ultimate explanation for the DE in
our Universe. It is also hard to acquiesce in the idea that we
accidentally happen to live in an epoch where the CC contributes an
energy density value $\,\rL=\CC/8\pi\,G\,$ right in the ballpark of
the rapidly diluting matter density $\rmr\sim 1/a^3$. It should
perhaps be more plausible to conceive that the vacuum energy, $\rL$,
is actually a dynamical quantity as the Universe itself. More
generally, we could even entertain the possibility that the total DE
is in fact a mixture of $\rL$ and other dynamical components (e.g.
fields, higher order terms in the effective action etc) which can be
represented collectively by an effective entity $X$ (dubbed the
``cosmon''). The ``cosmon'', therefore, acts as a dynamical DE
component different from the vacuum energy. While it can actually
behave phantom-like by itself, the overall DE fluid may effectively
appear as standard quintessence, or even mimic at present an almost
exact CC behavior. Thanks to the versatility of such cosmic fluid we
can show that a composite DE system of this sort (``$\CC$XCDM'') may
have a key to resolving the mysterious coincidence problem.

\end{abstract}
%\pacs{04.62.+v, 95.36.+x, 98.80.Cq}

\section{Introduction}

Modern Cosmology has reached a status of a mature empirical science.
It is still far away from the level of precision of Particle
Physics, but it is on the way. Independent data sets derived from
the observation of distant supernovae\,\cite{SNe}, the temperature
anisotropies of the CMB\,\cite{WMAP3Y}, the integrated Sachs-Wolfe
effect\,\cite{SW}, the lensing corrections on the propagation of
light through weak gravitational fields\,\cite{Lensing}, and the
inventory of cosmic matter from the large scale structures (LSS) of
the Universe\,\cite{LSS} indicate altogether that our Universe is
presently under a phase of accelerated expansion. It is of course
tempting to simplify this state of affairs by just resorting to the
existence of an absolutely constant (time-independent) CC term,
$\CC$, in Einstein's equations. It is no less tempting to supersede
this hypothesis with another -- radically different-- one: viz. to
introduce a slowly evolving scalar field $\phi$ (``quintessence'')
whose potential, $V(\phi)\gtrsim 0$, accounts for the present value
of the DE and whose equation of state (EOS) parameter
$\omega_{\phi}= p_{\phi}/\rho_{\phi}\simeq -1+\dot\phi^2/V(\phi)$ is
only slightly larger than $-1$ (hence insuring a negative pressure
mimicking the $\CC$ case). In this way the DE can be a dynamical
quantity taking different values throughout the history of the
Universe. However, this possibility can not easily explain why the
present value of the DE is so close to the rapidly decaying matter
density, $\rmr\sim 1/a^3$ -- the so-called ``cosmic coincidence
problem''. And even if it could (as some modified quintessence
models propose\,\cite{Amendola}), there is after all a vacuum energy
associated to the other fields (e.g. the electroweak Standard Model
ones) and, therefore, such hypothetical scalar field cannot be the
only one responsible for the vacuum energy. At the end of the day it
does not seem to be such a wonderful idea to invent a field $\phi$
and simply replace $\rL$ with $\rho_{\phi}\simeq V(\phi)$. More
fundamentally, $\CC$ could instead  be conceived as a ``running
parameter'' in QFT in curved space-time, as proposed
in\,\cite{JHEPCC1}. Here we go a bit beyond and suggest that the DE
could involve, apart from a dynamical $\CC$, another collective
component, ``$X$'', which does not necessarily represent any
\textit{ad hoc} field. It may stand for higher order terms in the
effective action, perhaps in combination with some low-energy
``relics'' (e.g. a dilaton) from string theory, but in any case
without being a full substitute for $\CC$. Of course we have to
assume that the corresponding energy densities $\rX$ and $\rL$
conspire so as to generate the tiny value of the DE density at
present -- the ``old CC problem''\,\cite{weinberg}. While we cannot
solve this problem at this stage, the dynamical nature of $\CC$ and
$X$ gives at least allowance for this possibility to occur.

Here we focus on the second CC problem, the ``coincidence
problem''\,\cite{weinberg}. As we shall see, in the present
framework we can provide a novel clue for a possible resolution of
this problem. To start with, we note that once we impose the Bianchi
identity in Einstein's equations (derived from the full effective
action) it acts as a kind of ``superselection energy sum rule''
whereby the many terms on the \textit{r.h.s} of these ``effective''
Einstein's equations must add up themselves to satisfy a local
energy conservation law, irrespective of the inner details of the
particular model. One of these terms is of course the vacuum energy,
$\rL$, and the other terms can be treated as the aforesaid single
effective entity ``$X$'', which we will refer to sometimes as the
``cosmon''. For obvious reasons we call this class of composite
$(\CC,X)$-dark energy models the ``$\CC$XCDM
models''\,\cite{GSS1,GSS2}. In them we still have some freedom in
the way matter, vacuum and cosmon energy densities realize the local
conservation law. Here we will explore just two possibilities that
we call ``type I'' and ``type II'' $\CC$XCDM models. In type I
models matter is conserved and the total dynamical DE
($\rD=\rL+\rX$) too. In type II models, instead, matter and cosmon
densities are separately conserved, but the $\CC$ variation is
compensated for by a variable gravitational coupling $G$.  In the
following we expand on these two possibilities and show that any of
them could efficiently solve the coincidence problem.

\section{Composite dark energy models}

\begin{figure}[t]
  %\begin{center}
    \begin{tabular}{cc}
      \resizebox{0.45\textwidth}{!}{\includegraphics{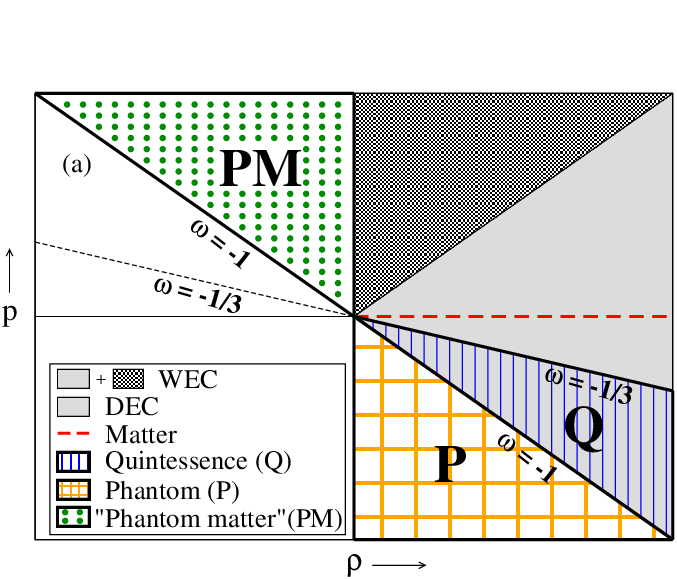}} &
      \hspace{0.3cm}
      \resizebox{0.45\textwidth}{!}{\includegraphics{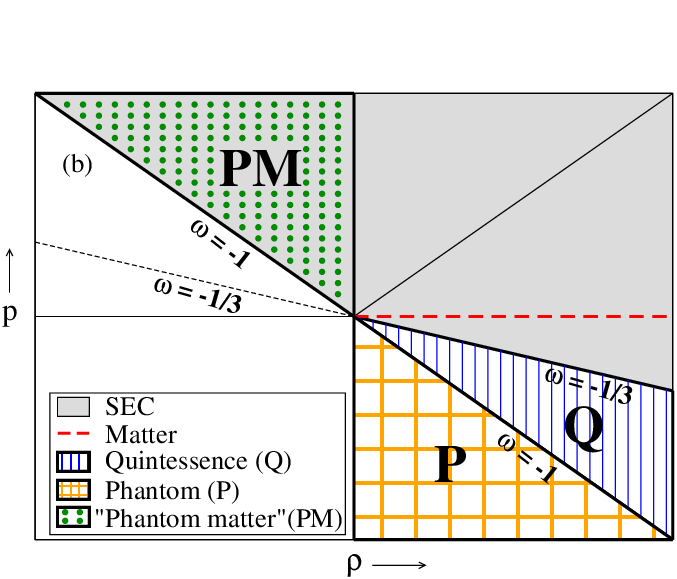}} \\
      %(a) & (b)
    \end{tabular}
\caption{\textbf{(a)} The Weak (WEC, $\rho\ge0$ and $\rho+p\ge0$)
and Dominant (DEC, $\rho\ge|p|$) energy conditions. The EOS regions
of matter, quintessence (Q) and ``standard'' phantom (P) are shown,
together with the ``phantom matter'' (PM) region ($\we<-1$ with
$\rho<0$); \textbf{(b)} The Strong energy condition (SEC,
$\rho+p\ge0$ and $\rho+3p\ge0$) insuring attractive gravitation
force, is only satisfied by matter and PM.} \label{fig1}
\end{figure}

For a DE medium consisting of several fluids with\,
$p_i=\omega_i\,\rho_i\ (i=1,2,...,n)$, the effective EOS parameter
of the mixture reads:
\begin{equation}\label{mix}
\we=\frac{\pD}{\rD}=\frac{\omega_1\,\rho_1+\omega_2\,\rho_2+...}{\rho_1+\rho_2+...}\,,
\end{equation}
being in general a function of time or the (cosmological) redshift
$z$, even if all $\omega_i$ are constant. Assuming a flat FLRW
metric, Einstein equations for such a model lead to:
\begin{eqnarray}
H^2=\frac{8\pi G}{3}\left(\rmr+\rho_1+\rho_2+...\right)\label{fri1}\\
\frac{\ddot{a}}{a}= -\frac{4\pi
G}{3}\left[\rmr(1+3\wm)+\rho_1(1+3\omega_1)+\rho_2(1+3\omega_2)...\right]ç,,
\label{acel}
\end{eqnarray}
where $\rmr$ stands for the density of matter-radiation. Equation
(\ref{fri1}) entails the following ``generalized cosmic sum rule''
valid at any $z$:
\begin{equation}\label{sumrule1}
\fl \Omh(z)+\hat{\Omega}_1(z)+\hat{\Omega}_2(z)+...=1\,;\ \
\hat{\Omega}_i(z)\equiv\frac{\rho_i(z)}{\rc(z)}=\frac{8\pi G
\rho_i(z)}{3H^2(z)}\ (i={\rm m},1,2,...)\,.
\end{equation}
We denote $\hat{\Omega}_i(z=0)\equiv{\Omega}_i^0$. If the DE is a
single fluid, the corresponding sum rule at present
($\Om^0+\OD^0=1$) enforces $\rDo$ to be positive, given that
$\Om^0\simeq0.3$\,\cite{SNe} . If in addition the DE is
self-conserved, then necessarily $\rD>0$ at any time. On the other
hand, for a composite DE it is clear that (\ref{sumrule1}) could be
fulfilled even if one or more of the DE components have negative
energy density, and then many possibilities open up. For instance,
from (\ref{acel}) we note that a component with $\omega_i<-1/3$
\textit{and} $\rho_i<0$ would decelerate the expansion instead of
accelerating it (cf. Fig.\,\ref{fig1}). In particular, phantom-like
($\omega_i<-1$) components with negative energy density ($\rho_i<0$)
also observe the strong energy condition (SEC, cf.
Fig.\,\ref{fig1}b) just the same as matter. In contrast, the
``standard'' phantom components $(\omega_i<-1,\rho_i>0)$ not only
violate \textit{all} the classical energy conditions but also may
produce an specially acute anti-gravitational effect which
eventually leads to a kind of singularity known as ``Big
Rip''\,\cite{Phantom} whereby all bound systems, without exception,
are eventually ripped off and hence destroyed. At variance with
these ugly prospects, phantom components with $\rho_i<0$ cause a
fast deceleration that may result into the future stopping and
subsequent reversal of the Universe expansion. This is rather
significant since, as we will see, the stopping of the expansion can
be linked to the solution of the coincidence problem. Thus, phantom
components with negative energy density (hence positive pressure)
behave as a sort of ``unclustering matter'' pervading the Universe;
it has ben called ``Phantom matter'' (cf. PM region in
Fig.\,\ref{fig1})\,\cite{GSS1}. The aforementioned ``cosmon'', for
instance, can behave as PM.

\section{$\CC$XCDM cosmologies}

As a simple realization of the idea of a composite DE we have
considered a dual DE system consisting of a running $\CC$ in
potential interaction with another dynamical entity $X$ (which we
have referred to as the ``cosmon'' -- see \cite{PSW} for the origin
of the name). As explained in Section 1, the $X$ component would
stand for any contribution to the DE energy other than the vacuum
energy effects. We call this scenario the $\CC$XCDM
model\,\cite{GSS1}. The evolution of $\CC$ is (as any parameter in
QFT) tied to the renormalization group (RG) in curved
space-time\,\cite{JHEPCC1}. The running equation we will consider
here for the CC density is one that has been thoroughly studied in
the literature\,\cite{JHEPCC1,RGTypeIa}:
\begin{equation}\label{RGEG1a}
\frac{\rmd\rL}{\rmd\ln \mu}=\frac{\sigma}{(4\pi)^2}
M^2\,\mu^2\equiv\frac{3\,\nu}{4\,\pi}\,\MP^2\,\mu^2\,.
\end{equation}
Here $\mu$ is the energy scale associated to the RG in Cosmology
(that we will identify with the Hubble function, i.e. $\mu=H$ at any
epoch, see\,\cite{RGTypeIa}) and
$\nu=({\sigma}/{12\pi}){M^2}/{\MP^2}$ is a free parameter, $M$ being
the effective mass of the heavy particles contributing to the
$\beta$-function of the CC and $\sigma=\pm 1$ depending on whether
bosons or fermions dominate. If $M=\MP$ (Planck mass), $|\nu|$ takes
the value $\nu_0\equiv 1/12\pi\simeq  0.026$. Thus, we naturally
expect $|\nu|\ll 1$. Introducing an specific model for $\CC$ allows
us to preserve the generality of the cosmon $X$. Let us only mention
that it could e.g. be some scalar field $\chi$ resulting from
low-energy string theory (e.g. a pseudo-dilaton, as in the original
paper \cite{PSW}), or account for the effective behavior of a
mixture of dynamical fields of various sorts and/or higher order
curvature terms in the action. We really do not need to specify its
ultimate nature here because, as we shall see, the kind of
cosmological implications that we will investigate (in particular
its impact on the coincidence problem ) do not depend on it.

We start the formulation of the model from the most general form of
the Bianchi identity on both sides of Einstein's equations:
$\nabla^{\nu}G_{\mu\nu}=8\pi\,\nabla^{\nu}(GT_{\mu\nu})=0$, where
$T_{\mu\nu}$ is the effective energy-momentum tensor including all
the terms on the \textit{r.h.s} of Einstein's equations. Assuming a
FLRW metric and describing $T_{\mu\nu}$ as a mixture of fluids
(including matter), the $\mu=0$ component of the Bianchi identity
above yields:
\begin{equation}\label{generalBD}
\frac{\rmd}{\rmd
t}\left[G\,\left(\sum_i\rho_i\right)\right]+G\,H\,\sum_i
\alpha_i\,\rho_i=0\,,\qquad\alpha_i\equiv 3(1+\omega_i)\,,
\end{equation}
where the  various $\rho_i$ stand for the energy densities of the
different fluids. In our case we have matter-radiation and a
composite DE:
\begin{eqnarray}
\rmr=\rM+\rR \,, \ \ \ \ \ \ \ \rD=\rX+\rL\,, \label{fosc}
\end{eqnarray}
$\rX(t)$ being the dynamical density of the cosmon and $\rL(t)$ the
energy density of the running $\CC$. In principle we also admit the
possibility of having a variable Newton's coupling, $G=G(t)$. Note
that this time variability of the cosmological parameters is
consistent with the Cosmological Principle and that the general
covariance of the theory is insured by the fulfilment of the general
Bianchi identity (\ref{generalBD}). The two DE components have EOS
parameters $\omega_{\CC}=-1$ and $\wX$ respectively, with:
\begin{equation}\label{wxrange}
-1-\delta/3<\wX<-1/3\Longrightarrow-\delta<\aX<2<\amr\qquad(\delta\gtrsim
0)\,
\end{equation}
i.e. $X$ can be quintessence or phantom-like. We consider two
possible realizations of the $\CC$XCDM  model, in both cases with
local conservation of matter:
\begin{enumerate}
\item Type I $\CC$XCDM model: $\dot{G}=0$ and $\rmr$ conserved,
hence $\rD$ is also conserved.
\item Type II $\CC$XCDM model: $\dot{G}\neq0$ with $\rmr$ and $\rX$
conserved.
\end{enumerate}
Next we will explain the main features of these two models, showing
that they can both alleviate the coincidence problem and match the
available data. Comparatively, modified (interactive) quintessence
models can tackle the coincidence problem only at the expense of
matter non-conservation\,\cite{Amendola}.  Herein we limit ourselves
to the simplest case, in which spatial flatness and constant $\wX$
are assumed, although more general results can be obtained
analytically \cite{GSS1}.

\section{Type I $\CC$XCDM models}

In this case we assume that Newton's coupling $G$ is constant and
matter-radiation is conserved, $\drmr+\amr\,\rmr\,H=0$. These
conditions in turn imply the conservation of the total DE density
$\rD$. The relevant set of equations are the Friedmann equation
(\ref{fri1}), the conservation of DE arising from the Bianchi
identity (\ref{generalBD}), and the RG model for $\CC$
(\ref{RGEG1a}):
\begin{eqnarray}
H^{2}=\frac{8\pi\,G }{3}\left( \rmr +\rL+\rX\right)\,,\label{FL}\\
\dot{\rho}_{\rm D}+\alpha_{\rm D}\,\rD\,H=\dot{\rho}_{\Lambda}+\drX+\,\aX\,\rX\,H=0\,,\label{B2}\\
\frac{\rmd\rL}{\rmd\ln
H}=\frac{3\,\nu}{4\,\pi}\,\MP^2\,H^2\,\label{RGEG1b}.
\end{eqnarray}
These equations can be rewritten as an autonomous system in terms of
the new independent variable $\zeta=-\ln (1+z)$ (the asymptotic past
and future lying at $\zeta=-\infty$ and $\zeta=\infty$ respectively
and the present at $\zeta=z=0$):
\begin{eqnarray}\label{autonomous1}
\OX'=-\left[\nu\,\amr+(1-\nu)\,\aX\right]\,\OX-\nu\,\amr\,\OL+\nu\,\amr\,\Oc\,,\nonumber\\
\OL'=\nu\,(\amr-\aX)\,\OX+\nu\,\amr\,\OL-\nu\,\amr\,\Oc\,,\\
\Oc'=(\amr-\aX)\,\OX+\,\amr\,\OL-\amr\,\Oc\,,\nonumber
\end{eqnarray}
where $\phantom{A}' \equiv \rmd/\rmd\zeta$. In the previous system
the density fractions ${\Omega}_i$ are normalized with respect to
the present critical energy density (in contrast to the
$\hat{\Omega}_i$ in (\ref{sumrule1})):
\begin{equation}\label{omedef}
\Omega_i(z)=\frac{\rho_i(z)}{\rco}=\frac{8\pi G \rho_i(z)}{3H_0^2}\
\qquad(i={\rm X},\CC,{\rm c})\,.
\end{equation}
Clearly, $\
\Omega_i(z)/\hat{\Omega}_i(z)=\,H^2(z)/H_0^2=\Omega_c(z)$. The
solution of (\ref{autonomous1}) reads as follows:
\begin{equation}\label{solve2}
{\bf\Omega}(\zeta)\equiv\big(\OX(\zeta),\OL(\zeta),\Oc(\zeta)\big)=
C_1\,{\bf v_1}\,e^{\lambda_1\,\zeta}+C_2\,{\bf
v_2}\,e^{\lambda_2\,\zeta}+C_3\,{\bf v_3}\,,
\end{equation}
with:
\begin{eqnarray}\label{eig2}
\fl\lambda_1=-\aX\,(1-\nu)\,,&\lambda_2=-\amr\,,&\lambda_3=0\,.\nonumber\\
\fl{\bf v_1}=\big(1-\nu,\nu,1\big)\,,\qquad&{\bf v_2}=\Big(\frac{-\nu\,\amr}{\amr-\aX},\nu,1\Big)\,,\qquad&{\bf v_3}=\big(0,1,1\big)\,.\\
\fl
C_1=1-C_2-C_3\,,&C_2=\frac{\Omo(\amr-\aX)}{\amr-\aX\,(1-\nu)}\,,&C_3=\frac{\OLo-\nu}{1-\nu}\,.
\end{eqnarray}
The coefficients $C_k$ result from the boundary conditions at
present: $\Omega_i(0)=\Omega_i^0=\hat{\Omega}_i(0)$. Assuming the
aforementioned prior $\Omo\simeq0.3$, our model contains three free
parameters: $\nu$, $\wX$ and $\OLo=\ODo-\OXo\simeq0.7-\OXo$.

\subsection{Nucleosynthesis and the coincidence problem}

As the expansion rate is sensitive to the amount of DE, we must
ensure that our model does not spoil the predictions of primordial
nucleosynthesis. Thus, we will ask the ratio between DE and matter
radiation densities,
\begin{equation}\label{rz}
r(z)\equiv\frac{\rD(z)}{\rmr(z)}\,,
\end{equation}
to be small enough at the nucleosynthesis epoch, say $|\rN\equiv
r(z=\zN\sim10^9)|\lesssim10\%$ (\cite{RGTypeIa,SS1}, see also
\cite{Ferreira97}). From the solution (\ref{solve2}) and after a
straightforward calculation we find that:
\begin{equation}\label{nb1}
|\rN|<10\%\,\Longleftrightarrow\,\frac{|\epsilon|}{\wR-\wX+\epsilon}\simeq|\epsilon|<0.1\,,\qquad\epsilon\equiv\nu\,(1+\wX)\,,
\end{equation}
where $\wR=1/3$ is the EOS parameter of radiation. Note that only
for $\nu=0$ would the DE density be vanishing at the nucleosynthesis
time; this shows that, in general, in the type I $\CC$XCDM model the
presence of DE takes place at all epochs of the evolution. Looking
again at (\ref{rz}) but this time at the matter dominated era, one
can show that $r$ can have -at most- one extremum at some $z=\ze$ in
the future\,\cite{GSS1}. We are interested in the case in which this
extremum is a maximum, since this will help solving the cosmic
coincidence problem. Indeed, if $r$ remains bounded from above -and
maybe even of order $1$- for the entire Universe lifetime, the fact
that $r_0\equiv r(z=0)\sim 1$ would no longer look like a
coincidence. The conditions for the extremum to exist and to be a
maximum are shown to be \cite{GSS1}:
\begin{equation}\label{maxcond}
\fl\frac{\OLo-\nu}{\wX\,(\OXo+\nu\,\OMo)-
\epsilon\,(1-\OLo)}>0\qquad\mbox{and}\qquad\aX\,\left(\OLo-\nu\right)<0\,.
\end{equation}

Remarkably, the existence of a maximum for $r(z)$ entails a halt of
the cosmic expansion at some future point $z=\zs$ (stopping
redshift). To prove this, let us notice that Einstein's equations
(\ref{FL}) and (\ref{acel}) for the type I $\CC$XCDM model imply:
\begin{eqnarray}
\lim_{z\rightarrow-1}H^2/H_0^2=\lim_{z\rightarrow-1}\OD\,,\label{FL-1}\\
\left.\frac{\ddot{a}}{a}\right|_{t=t_0}=-\frac{4\pi\,G}{3}\,[(1+r_0)+r'(0)]\,,\label{acc}
\end{eqnarray}
\noindent where $z\rightarrow -1$ stands for the remote future,
$t_0$ is the present time and $r'(z)=\rmd r(z)/\rmd z$. Since
$r_0>0$, the current state of accelerated expansion requires
$r'(z=0)<0$ (i.e. $r$ is presently increasing with time:
$\dot{r}(t=t_0)>0$). Now, if the \textit{r.h.s} of (\ref{FL-1}) is
positive, $\lim_{z\rightarrow-1}r(z)=\infty$ and the ratio is
unbounded. Moreover, knowing that the function $r=r(z)$ can have
\emph{at most} one extremum \cite{GSS1} the foregoing conditions
imply that there cannot be any extremum in the future. Thus, in this
case the DE cannot become negative to stop the expansion. If,
instead, the \textit{r.h.s} of (\ref{FL-1}) is negative, then
$H(\zs)=0$ at some $\zs>-1$. However, being $r$ presently positive
and increasing with time, this situation can only be compatible with
$\lim_{z\rightarrow -1}r(z)=-\infty$ if there is a maximum of $r$ at
some point between $z=0$ and $z=\zs$ (\textit{q.e.d.}). Note that
the condition $r'(0)<0$ and the uniqueness of the extremum enforce
the maximum to occur always in the future.

\subsection{Some possible scenarios}

Within the type I $\CC$XCDM model there are many scenarios
compatible with stopping (and subsequent reversal) of the expansion
in the future. As shown in the last section, this implies the
existence of a maximum of the ratio $r$ with the consequent
alleviation of the coincidence problem. A very convenient form to
identify these scenarios is by studying the phase trajectories of
the autonomous system (\ref{autonomous1}). Notice that an
asymptotically negative value of the third component of
(\ref{solve2}), $\Oc(z)\equiv H^2(z)/H_0^2$, would indicate the
existence of a stopping point. Let us discuss here just two
representative cases -- for a comprehensive analysis
see\,\cite{GSS1}:

\begin{itemize}

\item 1) $-\delta<\aX<0$ (phantom-like cosmon) and $\nu<1$.
Looking at the eigenvalues (\ref{eig2}), we see that
$\lu>0\,,\ld<0$, so there is a saddle point in the phase space:
\begin{equation}
{\bf \Omega}^{*}=(0,\,\OLo,\,\OLo)
\end{equation}
from which all trajectories diverge with the evolution (as
$\zeta\rightarrow\infty$). However, if $C_1<0$ in (\ref{eig2}),
$\Oc(z\rightarrow-1)<0$, implying the stopping of the expansion as
discussed above. Using (\ref{nb1}), the stopping condition acquires
the form:
\begin{equation}\label{C1approx}
\fl
C_1=1-C_2-C_3=\frac{1-\OLo}{1-\nu}-\frac{\OMo(\wm-\wX)}{\wm-\wX+\epsilon}\simeq\frac{1-\OLo}{1-\nu}-\OMo<0\,.
\end{equation}
We can check that this relation insures: a) the fulfilment of the
conditions (\ref{maxcond}) for the existence of a maximum; b)
$\OXo<-\nu\,\OMo$. Thus for $0<\nu<1$ the cosmon behaves as PM,
whilst for $\nu<0$ behaves as a ``conventional'' phantom.

\item 2) $0<\aX<2$ (quintessence-like cosmon) and $\nu<1$. Now the non-vanishing eigenvalues are both negative
$\lu<0\,,\ld<0$. Therefore, there is a ($\nu$-dependent) node
towards which all phase trajectories are attracted, namely
\begin{equation}\label{nunode}
{\bf\Omega}^{*}=\left(0,\,\frac{\OLo-\nu}{1-\nu},\,\frac{\OLo-\nu}{1-\nu}\right)\,.
\end{equation}
This time, the attraction towards the node will be stopped for the
curves satisfying:
\begin{equation}\label{cond1}
\frac{\OLo-\nu}{1-\nu}<0\qquad\Rightarrow\qquad\OLo<\nu<1\,.
\end{equation}
\end{itemize}

\begin{figure}[t]
  %\begin{center}
    \begin{tabular}{cc}\\
      \resizebox{0.45\textwidth}{!}{\includegraphics{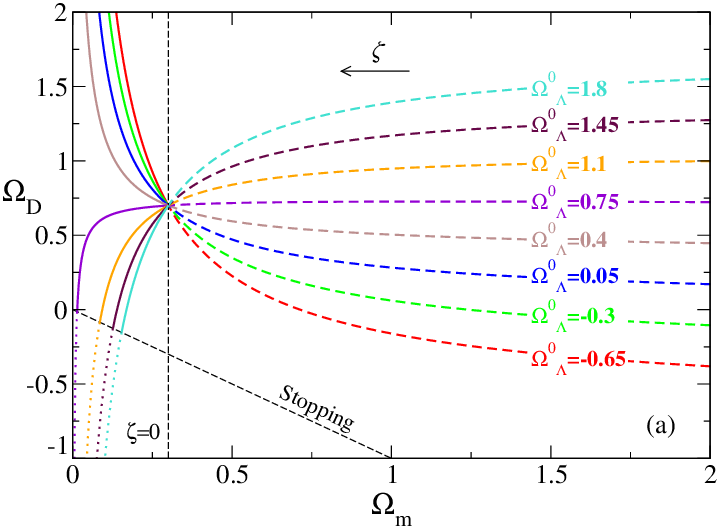}} &
      \hspace{0.3cm}
      \resizebox{0.45\textwidth}{!}{\includegraphics{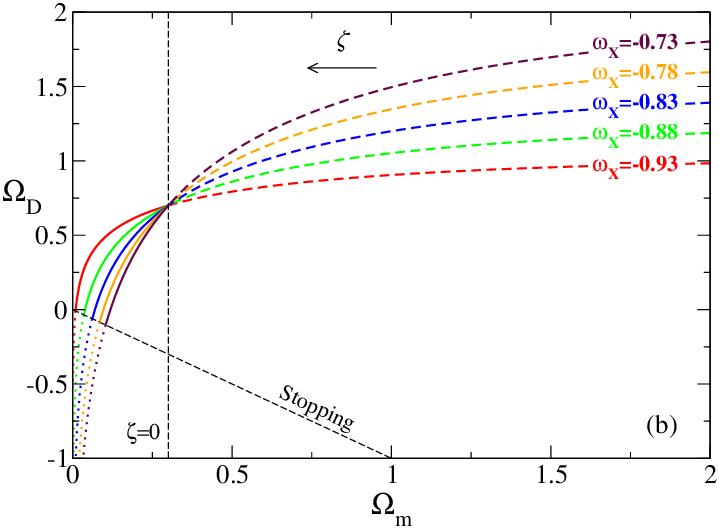}} \\
     % (a) & (b)
    \end{tabular}
\caption{Phase trajectories of the autonomous system
(\ref{autonomous1}) in the $(\Om,\OD)$ plane for different values of
the parameters. Dashed lines show the parts of the curves
corresponding to our past, full lines the parts between the present
moment and the stopping (if there is stopping) and dotted lines the
inaccessible part of the trajectory after the stopping; \textbf{(a)}
$\wX=-1.85$, $\nu=-\nu_0$ (cf. case 1 in the text) and different
choices of $\OLo$ \textbf{(b)} $\OLo=-2$, $\nu=0.96$ (cf. case 2)
and different values of $\wX$.}
  \label{fig2}
\end{figure}

The projections onto the ($\Om,\OD$) plane of the phase trajectories
for these two scenarios have been plotted in Fig.\,\ref{fig2}a,b.
They show respectively the existence of a saddle point or a node and
the stopping of the curves that fulfil (\ref{C1approx}) or
(\ref{cond1}).
Note that the solution of the coincidence problem can take place in
both scenarios even for the simplest situation, namely for $\nu=0$.
In this case, the DE of the model is just the sum of a
self-conserved cosmon and a strictly constant $\CC$:
\begin{equation}\label{ODznu0}
\OD(z)=\OLo+ \OXo\,(1+z)^{\aX}\,,
\end{equation}
and the stopping conditions deduced above just read: $\OXo<0$ (case
1) and $\OLo<0$ (case 2), corresponding to a Universe containing
phantom matter or a negative $\CC$ respectively. In both cases the
overall effective EOS of the DE is quintessence-like. Even though
$\nu=0$ does the job as far as the cosmic coincidence problem is
concerned, the possibility to solve that problem also for $\nu\neq
0$ allows us to modulate the effective EOS behavior of the model and
describe features that can be potentially observed in it (see next
section).

\subsection{A numerical example}

\begin{figure}[t]
  %\begin{center}
    \begin{tabular}{cc}\\
      \resizebox{0.47\textwidth}{!}{\includegraphics{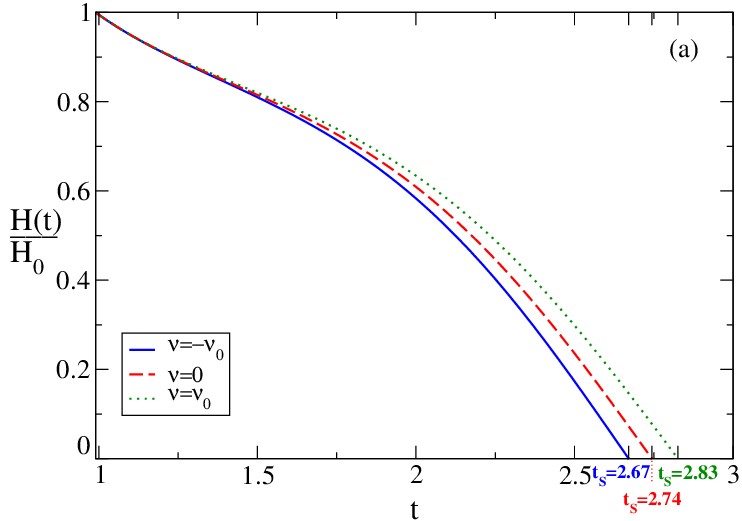}} &
      \hspace{0.3cm}
      \resizebox{0.47\textwidth}{!}{\includegraphics{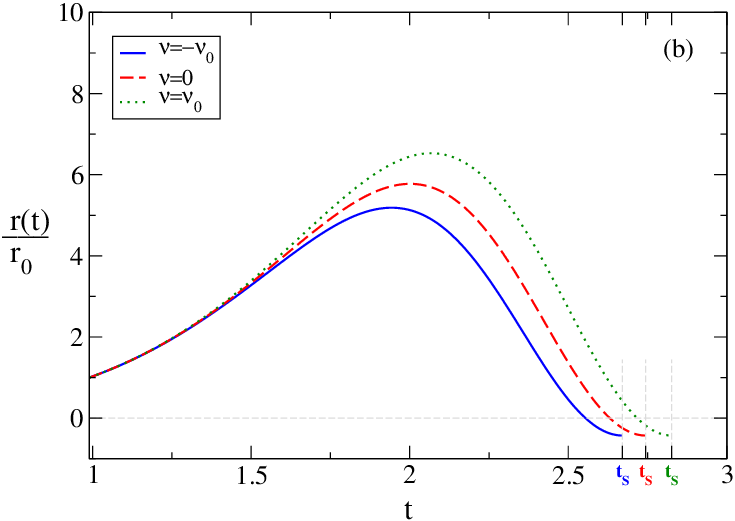}} \\
     % (a) & (b)
    \end{tabular}
\caption{\textbf{(a)} The expansion rate (normalized to its present
value $H_0$) as a function of the cosmic time, $t$, in the type I
model for the following parameter values: $\wX=-1.85$, $\OLo=0.75$
and $\nu=-\nu_0,\,0,\,+\nu_0$, showing the existence of a stopping
point; \textbf{(b)} The ratio $r=\rD/\rmr$ (in units of its present
value $r_0$) as a function of $t$, illustrating the presence of both
a maximum and a stopping point in the future, what constitutes a
possible explanation of the coincidence problem. The cosmic time $t$
is measured in Hubble time units $H_0^{-1}$; the present moment lies
at $t\simeq 0.99$ (i.e. $13.7\,Gyr$).}
  \label{fig3}
\end{figure}

\begin{figure}[t]
  %\begin{center}
    \begin{tabular}{cc}\\
      \resizebox{0.45\textwidth}{!}{\includegraphics{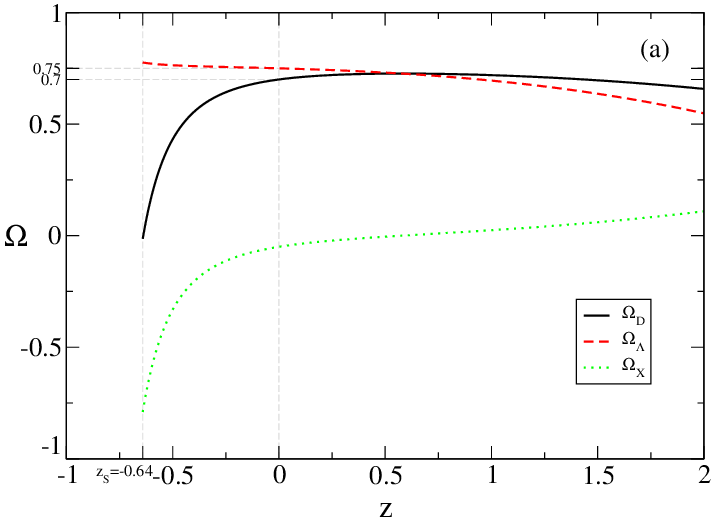}}
      \hspace{0.3cm}
      \resizebox{0.45\textwidth}{!}{\includegraphics{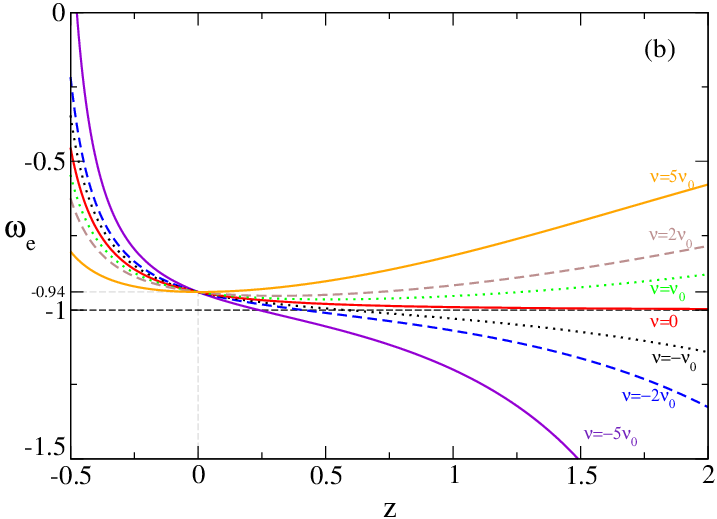}} \\
      %(a) & (b)
    \end{tabular}
\caption{\textbf{(a)} The evolution of the density fraction $\Omega$
for the total DE and each one of its components, $X$ and $\CC$, in
the type I model. The values of the parameters are the same as in
Fig.\,\ref{fig3} with the choice $\nu=-\nu_0$. We see that the
stopping of the expansion (characterized by the value
$\OD(z_s)=-\Om(z_s)$) is achieved thanks to $\OX$ being negative,
i.e. to the behavior of the cosmon as PM; \textbf{(b)} Different
behaviors of the EOS of the $\CC$XCDM model for the same $\wX$ and
$\OLo$ as in Figure\,\ref{fig3} and different values of $\nu$.}
  \label{fig4}
\end{figure}

Let us now illustrate the properties of the model through an
specific example, namely $\wX=-1.85$, $\OLo=0.75$ and different
values of $\nu<1$. These values correspond to the first scenario
discussed in the previous section. As long as we take $\nu$
sufficiently small, condition (\ref{C1approx}) will be satisfied and
we should get both the stopping of the expansion and a maximum in
the ratio $r$. In Fig.\, \ref{fig3}a,b we can observe these features
for three different values of $\nu$ ($0$ and $\pm\nu_0$). There we
have plotted the Hubble function and the ratio $r$ as functions of
the cosmic time -- which requires solving numerically the model. We
see that the maximum and the stopping point take place far away in
the future. In these cases, the halt of the expansion is caused by
the behavior of the cosmon as PM rather than to a negative $\CC$, as
we can appreciate in Fig.\,\ref{fig4}a. In this plot we also notice
that the signs of the two components of the DE (which are
individually unobservable) can change during the evolution.
\\
Of course, in order to claim that our model provides a solution, or
at least an important alleviation of the coincidence problem, we
must be sure that this holds for a significant part of the full
parameter space. We have confirmed this point through a
comprehensive numerical sampling of it, and moreover we have found
that the maximum of the ratio $r$ can stay rather small, say $1-10$
times its present value $r_0$ \cite{GSS1}.

The comparison between a particular DE model and the observational
data is made mainly through the EOS. The effective EOS of the type I
$\CC$XCDM model is given by:
\begin{equation}\label{eEOS}
\we^{(I)}=\frac{\pD}{\rD}=\frac{\pL+\pX}{\rL+\rX}=\frac{-\rL+\wX\,\rX}{\rL+\rX}=
-1+(1+\wX)\,\frac{\rX}{\rD}\,,
\end{equation}
and it can display a rich variety of behaviors while being in
agreement with the most recent data\,\cite{WMAP3Y}. This is shown in
Fig.\,\ref{fig4}b, where we see that, depending on the value of
$\nu$, the EOS can be quintessence-like, mimic (in some cases almost
exactly) a pure CC term or even present a mild transition from the
phantom to the quintessence regime.

\subsection{Asymptotic behavior in the past: a signature of the model}

Let us finally elaborate on a characteristic feature of type I
models with running $\CC$ that could serve to distinguish them from
other DE models. From the solution (\ref{solve2}), we find that in
the asymptotic past:
\begin{eqnarray}
\OD(z\gg 1)=-\frac{\epsilon}{\wm-\wX+\epsilon}\,\Omo\,(1+z)^{\amr}\,,&(\nu\neq0)\\
\we^{(I)}(z\gg 1)=-1+(1+\wX)\,\frac{\OX(z\gg 1)}{\OD(z\gg
1)}\rightarrow\wm\,,\qquad&(\nu\neq 0)\,.
\end{eqnarray}
Thus, surprisingly, at very high redshift the effective EOS of the
DE coincides with that of matter-radiation:
$\we^{(I)}\rightarrow\wm$. Moreover, the Hubble function at high
redshift reads:
\begin{equation}\label{renormalization}
\fl H^2(z\gg1)\simeq H_0^2\,\Om^{0*}\,(1+z)^{\amr}\,,\qquad
\Om^{0*}=\Omo\,\left(1-\frac{\epsilon}{\wm-\wX+\epsilon}\right)\,,
\end{equation}
which means that the value  $\Om^{0*}$ of the density fraction of
matter inferred from very high $z$ data (e.g. from CMB) could differ
from that obtained by low $z$ experiments ($\Omo$), typically from
supernovae. The relative difference, $|\Om^{0*}-\Omo|/\Omo$, is
given just by the same expression as the nucleosynthesis constraint
(\ref{nb1}); therefore this effect could amount to a measurable
$10\%$ discrepancy and, if detected, would constitute a distinctive
signature of type I $\CC$XCDM models.

\section{Type II $\CC$XCDM models}

In this case we let the Newton coupling $G$ be variable and ask for
the conservation of matter-radiation. These two conditions allow in
turn the conservation of the $X$ component, $\drX+\aX\,\rX\,H=0$. We
adopt the same RG equation (\ref{RGEG1a}) as before for the running
of $\CC$. With all this in mind, we obtain the set of equations
defining the type II models:
\begin{eqnarray}\label{consLX2}
H^2=\frac{8\pi\,G}{3}\left(\rmr+\rL+\rX\right)\,,\\
(\rmr+\rL+\rX)\,dG+G\,d\rL=0\,,\\
\frac{\rmd\rL}{\rmd\ln \mu}=\frac{3\,\nu}{4\,\pi}\,\MP^2\,\mu^2\,.
\end{eqnarray}
This system can be analytically solved to determine $G$ as a
function of the scale $\mu=H$, with the following result:
\begin{equation}\label{GH}
G(H)=\frac{G_0}{1+\nu\,\ln\frac{H^2}{H_0^2}}\,,
\end{equation}
where $G_0=1/\MP^2$. This equation suggests that $\nu$ acts also as
the $\beta$-function for the RG running of $G$. We can also express
$G=G(z)$ as an implicit function of the redshift $z$:
\begin{equation}\label{implicit}
\fl\frac{1}{g(z)}-1+\nu\,\ln\left(\frac{1}{g(z)}-\nu\right)=
\nu\,\ln\left[\Omo\,(1+z)^{\amr}+\OXo\,(1+z)^{\aX}+\OLo
-\nu\,\right]\,,
\end{equation}
where we have defined $g(z)\equiv{G(z)}/{G_0}$. Once this function
is determined, $\rL=\rL(z)$ can also be derived:
\begin{equation}\label{CCz}
\rL(z)=\frac{\rLo+\nu\,\left(\rmr(z)+\rX(z)\right)\,g(z)-\nu\,\rc^0}{1-\nu\,\,g(z)}\,.
\end{equation}
Even though the model could be solved from these equations, it is
very convenient to find an effective equation of state approach to
it, since this will allow us to better compare our model to
alternative ones -e.g. quintessence scalar field models- and to
confront it with the experimentally measured EOS. The procedure is
thoroughly described in \cite{SS1,SS2}. The basic idea is that even
a model with a non-conserved DE (like this one) can be represented
by means of an ``effective EOS'' corresponding to a self-conserved
dark energy. Thus we can contemplate the type II model under two
different perspectives:
\begin{itemize}
\item In the \textit{original picture} we see the model as it is:
in this case, with self-conserved cosmon and matter-radiation
densities together with variable cosmological parameters $G$
(Newton's coupling) and $\rL$ (vacuum energy density).
\item In the \textit{DE picture} we assume that the model (regardless of its real
characteristics) has a constant $G$ and conserved DE and
matter-radiation. The effective EOS arising from this picture is the
one directly comparable to the experimental one.
\end{itemize}

The formula for $\we$ for type II models can be easily obtained from
the condition that the DE is conserved in the DE picture,
$\rmd{\tilde\rho}_{\rm D}/\rmd t+\alpha_e\,\rDt\,H=0$, where $\rDt$
is the DE density in this picture. It reads:
\begin{equation}
\we^{(II)}(z)=-1+\frac{1+z}{3\rDt(z)}\,\frac{\rmd\rDt(z)}{\rmd
z}\,.\label{EOSpar}
\end{equation}
As the expansion history should not depend on the picture, we can
match Hubble functions in each picture, hence:
$G(\rmr+\rL+\rX)=G_0\,(\rmr+\rDt)$. Substituting this equation in
(\ref{consLX2}) and solving for $\rmd\rDt(z)/\rmd z$ we obtain the
effective EOS parameter (\ref{EOSpar}) of the type II $\CC$XCDM
model in a suitable form:
\begin{equation}\label{EOS1}
\we^{(II)}(z)=-1+\frac{\delta(z)}{3\,\tOD(z)}\,,
\end{equation}
where
\begin{equation}\label{EOS2}
\delta(z)=\amr\left(g(z)\,\Omo-\tOmo\right)\,(1+z)^{\amr}+\aX\,g(z)\,\OXo\,(1+z)^{\aX}\,.
\end{equation}
Note that the various $\Omega$'s could depend (slightly) on the
picture as they result from different fits to the same data, i.e. in
general $\Delta\Omo\equiv\Omo-\tOmo\neq 0$. By requiring the same
nucleosynthesis condition as in the previous model (i.e.
$|r(\zN)|\lesssim 10\%$), we find that the only parameter
constrained this time is $\nu$, which must satisfy the rather severe
bound $|\nu|\lesssim 10^{-3}$ in order not to disturb
nucleosynthesis predictions. Recall that for type I models the bound
is on the parameter $\epsilon$ and is not so stringent, see
Eq.\,(\ref{nb1}).

\subsection{Numerical analysis of the EOS}

\begin{figure}[t]
  %\begin{center}
    \begin{tabular}{cc}
      \resizebox{0.45\textwidth}{!}{\includegraphics{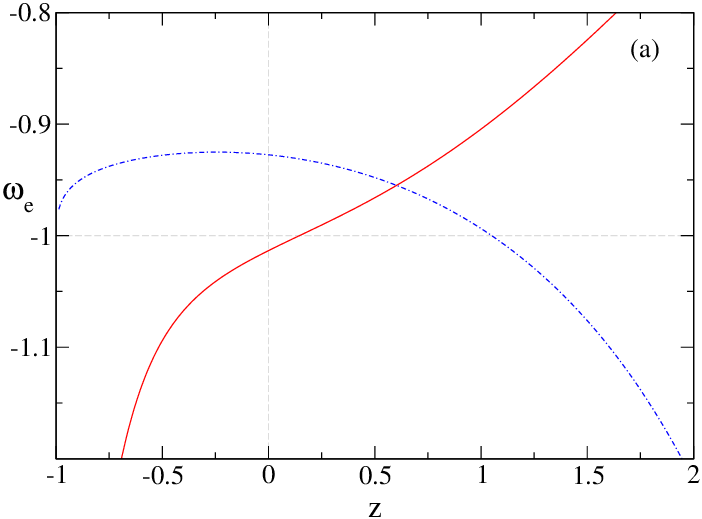}} &
      \hspace{0.3cm}
      \resizebox{0.45\textwidth}{!}{\includegraphics{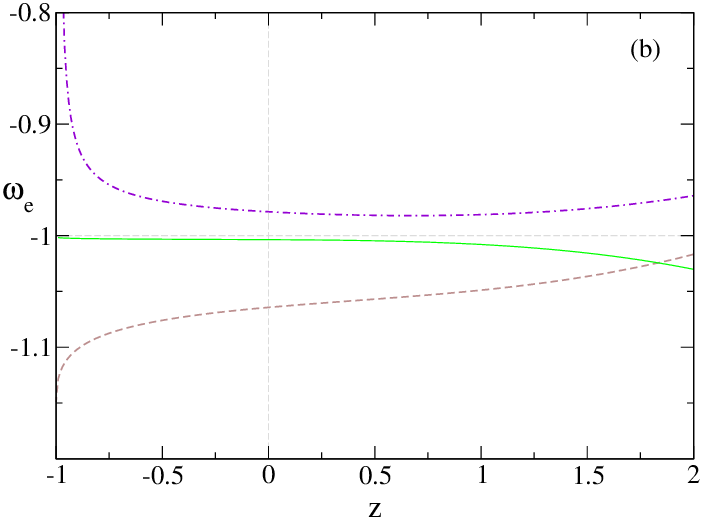}} \\
      %(a) & (b)
    \end{tabular}
\caption{Some examples of the behavior of the effective EOS function
for the type II cosmon models (\ref{EOS1}): \textbf{(a)}
$\wX=-1.65$, $\nu=+0.001$, $\OLo=0.67$ $\Delta\Omo=0.01$ (solid
line) and  $\wX=-0.85$, $\nu=-0.001$, $\OLo=0.3$, $\Delta\Omo=-0.01$
(dash-dotted), illustrating the two types of possible transitions in
the recent past; \textbf{(b)} $\wX=-0.95$, $\nu=0.001$, $\OLo=0.75$,
$\Delta\Omo=0$ (solid line) and $\wX=-1.15$, $\nu=-0.001$,
$\Delta\Omo=0$ and $\OLo=0.4$ (dashed) or $\OLo=0.8$ (dash-dotted),
showing examples in which the EOS parameter mimics that of a CC or
remains in the phantom or quintessence regime for all the redshifts
attainable by present and scheduled supernovae experiments.}
  \label{fig5}
\end{figure}

Although the coincidence problem can be solved similarly within the
type II $\CC$XCDM models\,\cite{GSS2}, here we will just focus on
the distinct behaviors of the effective EOS parameter (\ref{EOS1}),
which are exemplified in Fig.\,\,\ref{fig5}a,b. We see that we can
have transitions from quintessence into phantom regime (and vice
versa), or just remain quintessence or phantom-like across the
entire redshift interval relevant to SNIa observations. The model
can even closely mimic a pure CC term while retaining its dynamical
nature, which is revealed in its future behavior through the
presence of a stopping point. In all these cases, the current value
of $\we$ is close to $-1$, hence compatible with the available data
\cite{WMAP3Y}.
Finally, let us mention that in this model the behavior of $\we$ is
modulated by the parameter $\Delta\Omo$, which in this sense plays a
similar role as $\nu$ in the type I models.

\section{Conclusions}

$\CC$XCDM models with local conservation of matter reveal themselves
as generically capable to solve the cosmic coincidence problem, or
at least to highly mitigate it. We have shown that irrespective of
how the Bianchi identitity is implemented among the two density
components $(\rL, \rX)$ of the DE fluid, the overall behavior of the
cosmological model is such that the ratio between the total DE and
matter densities, $r=\rD/\rmr=(\rL+\rX)/\rmr$, remains bounded
throughout the entire history of the Universe, and can be of order
one at present. This result is model independent, in the sense that
we have not compromised the nature of the ``cosmon'' entity $X$ (in
particular we did not tie it to a scalar field with some peculiar
potential). We just used two independent ways of realizing the
geometric Bianchi identity through the local conservation laws of
the DE components, together with the local conservation of matter
and a renormalization group inspired law for the running of the
cosmological term.

\textit{Acknowledgements}. This work has been supported in part by
MEC and FEDER under project 2004-04582-C02-01, and also by DURSI
Generalitat de Catalunya under project 2005SGR00564. The work of HS
is also financed by the Ministry of Science, Education and Sport of
the Republic of Croatia.

%%%%%%%%%%%%%%%%%%%%%%%%%%%%%%%%%%%%%%%%%%%%%%%%%%%%%%%%%%%%%%%%%%%%%%%%%
%\newcommand{\JHEP}[3]{{\sl J. of High Energy Physics } {JHEP} {#1} (#2)  {#3}}
%\newcommand{\NPB}[3]{{\sl Nucl. Phys. } {\bf B#1} (#2)  {#3}}
%\newcommand{\NPPS}[3]{{\sl Nucl. Phys. Proc. Supp. } {\bf #1} (#2)  {#3}}
%\newcommand{\PRD}[3]{{\sl Phys. Rev. } {\bf D#1} (#2)   {#3}}
%\newcommand{\PLB}[3]{{\sl Phys. Lett. } {\bf #1B} (#2)  {#3}}
%\newcommand{\EPJ}[3]{{\sl Eur. Phys. J } {\bf C#1} (#2)  {#3}}
%\newcommand{\PRep}[3]{{\sl Phys. Rep } {\bf #1} (#2)  {#3}}
%\newcommand{\IJMP}[3]{{\sl Int. J. of Mod. Phys. } {\bf #1} (#2)  {#3}}
%\newcommand{\PRvL}[3]{{\sl Phys. Rev. Lett. } {\bf #1} (#2) {#3}}
%\newcommand{\ZFP}[3]{{\sl Zeitsch. f. Physik } {\bf C#1} (#2)  {#3}}
%\newcommand{\MPLA}[3]{{\sl Mod. Phys. Lett. } {\bf A#1} (#2) {#3}}
%%%%%%%%%%%%%%%%%%%%%%%%%%%%%%%%%%%%%%%%%%%%%%%%%%%%%%%%%%%%%%%%%%%%%%%%%%
%%%%%%%%%%%%%%%%%%%%%%%%%%%%%%%%%%%%%%%%%%%%%%%%%%%%%%%%%%%%%%%%%%%%%%%%%%
%\newcommand{\CQG}[3]{{\sl Class. Quant. Grav. } {\bf #1} (#2) {#3}}
%\newcommand{\JCAP}[3]{{\sl J. of Cosmology and Astrop. Phys. }{ JCAP} {\bf#1} (#2)  {#3}}
%\newcommand{\APJ}[3]{{\sl Astrophys. J. } {\bf #1} (#2)  {#3}}
%\newcommand{\AMJ}[3]{{\sl Astronom. J. } {\bf #1} (#2)  {#3}}
%\newcommand{\APP}[3]{{\sl Astropart. Phys. } {\bf #1} (#2)  {#3}}
%\newcommand{\AAP}[3]{{\sl Astron. Astrophys. } {\bf #1} (#2)  {#3}}
%\newcommand{\MNRAS}[3]{{\sl Mon. Not.Roy. Astron. Soc.} {\bf #1} (#2)  {#3}}
%%%%%%%%%%%%%%%%%%%%%%%%%%%%%%%%%%%%%%%%%%%%%%%%%%%%%%%%%%%%%%%%%%%%%%%%%

\end{document}